\documentclass[journal,twoside]{IEEEtran}
\usepackage{amssymb}
\usepackage{amsthm}
\usepackage{multirow}
\usepackage{inputenc}
\usepackage{enumerate} 
\usepackage{textcomp}
\usepackage{listings}
\usepackage{array}
\usepackage[switch,mathlines]{lineno}
\usepackage{soul}
\usepackage{xfrac}
\usepackage{graphicx}
\usepackage{cite}
\usepackage[cmex10]{amsmath}
\usepackage{algorithm}
\usepackage{xcolor}
\usepackage{colortbl}
\usepackage{cancel}

\makeatletter
\def\footnoterule{\kern-3\p@
 \hrule \@width 3.3in \kern 2.6\p@} 
\makeatother

\usepackage{algpseudocode}

\usepackage{stackengine}

\usepackage{algpseudocode}
\usepackage{CJK}
\usepackage[cmex10]{amsmath}
\usepackage{bm}
\usepackage{color}
\usepackage{amsmath,amsthm,amssymb,amsfonts}
\usepackage{flushend}
\usepackage{float}

\usepackage{arydshln} 
\usepackage{hyperref}
\hypersetup{
 colorlinks = true,
 linkcolor = blue,
 citecolor = red,
 urlcolor = black
}
\makeatletter
\newcommand*{\transpose}{%
 {\mathpalette\@transpose{}}%
}
\newcommand*{\@transpose}[2]{%
 \raisebox{\depth}{$\m@th#1\intercal$}%
}
\makeatother

\usepackage{soul}
\usepackage{tikz}
\usepackage{booktabs}
\usepackage{tabularx}

\usepackage[caption=false,font=footnotesize]{subfig}
\ifCLASSINFOpdf
\else
\fi

\begin{document}
\renewcommand{\ttdefault}{cmtt}
\bstctlcite{IEEEexample:BSTcontrol}

\title{Propagating Parameter Uncertainty in Power System Nonlinear Dynamic Simulations Using a Koopman Operator-Based Surrogate Model}

\author{
{Yijun~Xu,~\IEEEmembership{Senior Member}, 
Marcos~Netto,~\IEEEmembership{Member},
Lamine~Mili,~\IEEEmembership{Life Fellow}
}
\thanks{This work was authored in part by the National Renewable Energy Laboratory (NREL), operated by Alliance for Sustainable Energy, LLC, for the U.S. Department of Energy (DOE) under contract no. DE-AC36-08GO28308. This work was supported by the Laboratory Directed Research and Development (LDRD) program at NREL. The views expressed in the article do not necessarily represent the views of the DOE or the U.S. Government. The U.S. Government and the publisher, by accepting the article for publication, acknowledges that the U.S. Government retains a nonexclusive, paid-up, irrevocable, worldwide license to publish or reproduce the published form of this work, or allow others to do so, for U.S. Government purposes.}
\thanks{Y. Xu, and L. Mili are with the Bradley Department of Electrical and Computer Engineering, Virginia Tech, Northern Virginia Center, Falls Church, VA 22043, USA. M. Netto is with the Power Systems Engineering Center, NREL, Golden, CO 80401, USA. Corresponding author: \href{mailto:yijunxu@vt.edu}{yijunxu@vt.edu}.}
}


\maketitle

\begin{abstract}
We propose a Koopman operator-based surrogate model for propagating parameter uncertainties in power system nonlinear dynamic simulations. {\color{black}First, we augment a priori known state-space model by reformulating parameters deemed uncertain as pseudo-state variables. Then, we apply the Koopman operator theory to the resulting state-space model and obtain a linear dynamical system model. This transformation allows us to analyze the evolution of the system dynamics through its Koopman eigenfunctions, eigenvalues, and modes. Of particular importance for this letter, the obtained linear dynamical system is a surrogate that enables the evaluation of parameter uncertainties by simply perturbing the initial conditions of the Koopman eigenfunctions associated with the pseudo-state variables.} Simulations carried out on the New England test system reveal the excellent performance of the proposed method in terms of accuracy and computational efficiency.
\end{abstract}

\begin{IEEEkeywords}
Koopman operator; parameter uncertainty; statistical dynamic simulation; uncertainty propagation.
\end{IEEEkeywords}

\IEEEpeerreviewmaketitle
\vspace{-0.2cm}
\section{Introduction}
\IEEEPARstart{T}he uncertainties {\color{black}associated with electricity demand and supply, weather forecasting, measurement systems errors, and modeling accuracy} bring grand challenges to the design and operation of modern power systems. Thus, uncertainty quantification (UQ) has driven substantial research within the power system community. See, e.g., \cite{hiskens2006sensitivity, choi2016propagating, xu2018propagating}. In particular, propagating uncertainties in power system nonlinear dynamic simulations is an important problem and the focus of this letter.

Monte Carlo (MC) simulation is arguably the prevailing method for uncertainty propagation. Though straightforward, MC simulation exhibits a prohibitive computational burden for practical applications in sizeable electric power systems. Analytical methods based on a linear approximation \cite{hiskens2006sensitivity} of a nonlinear system model improve the computational efficiency of the simulations but at the expense of a significant loss of accuracy when the simulations involve events that push the system far from the system stable equilibrium point. Likewise, second-order approximations \cite{choi2016propagating} improve the accuracy but lose computational efficiency because they require the numerical evaluation of higher-order derivatives. Conversely, statistical methods \cite{xu2018propagating} simplify the approximation procedure while maintaining high computational efficiency but often lack physical meaning and interpretability. This letter proposes an alternative approach to propagate parameter uncertainty in power system nonlinear dynamic simulations based on the Koopman operator. Unlike analytical methods that perform first-order or second-order approximations of the system nonlinear model, the Koopman operator-based surrogate model captures the full nonlinear dynamics and is derivative-free. Unlike statistical methods, the proposed method retains physical interpretability and therefore is suitable for applications such as coherency identification \cite{susuki2011nonlinear} and selective modal analysis \cite{netto2018data}, among others. Furthermore, a Koopman operator-based surrogate \cite{lehmberg2020exploring, Peitz2020} of a power system nonlinear dynamic model enables the evaluation of a large set of parameters with low computational cost and high accuracy while propagating parameter uncertainties in power system dynamic simulations.

\vspace{-0.3cm}
\section{Koopman Operator}\label{section.II}
Let an autonomous nonlinear dynamical system evolving on a finite-dimensional manifold $M$ be governed by
\vspace{-.1cm}
\begin{equation}\label{eq.1}
\bm{\dot{x}}(t) = \bm{f}(\bm{x}(t)),
\end{equation}

\vspace{-.1cm}
\noindent
where $t\in\mathbb{R}$, $\bm{x}\in\mathbb{R}^{n_{x}}\subset{M}$ is the state, and $\bm{f}:M\to{M}$ is a nonlinear function. Let an observable $g(\bm{x})$ be a continuous function defined in $M$, $g:M\to\mathbb{R}$. The \emph{Koopman operator}, $\mathcal{K}_{t}$, is a linear, infinite-dimensional operator that acts on $g$,
\vspace{-.1cm}
\begin{equation}\label{koopman}
\mathcal{K}_{t}\,{g} = g (\bm{S}_{t}),
\end{equation}

\vspace{-.1cm}
\noindent
where $\bm{S}_{t}:{\color{black}M}\to{\color{black}M};\,\bm{x}(0)\to\bm{x}(t)=\bm{x}(0)+\int_{0}^{t}\bm{f}(\bm{x}(\tau))d\tau$ is called the flow. Because the Koopman operator is linear, its eigenvalues, $\lambda_{i}$, and eigenfunctions, $\phi_{i}$, are defined by
{\color{black}$\mathcal{K}_{t}\phi_{i}=e^{\lambda_{i}t}\phi_{i}$, $i=1,...,\infty$. In practice, one estimates a subset of the Koopman eigenvalues and eigenfunctions. To this end,} let $\bm{g}:{\color{black}M}\to\mathbb{R}^{n_{d}}$, $n_{d}\ge{n_{x}}$. If all $n_{d}$ elements of $\bm{g}$ lie within the span of the eigenfunctions $\phi_{i}$, then
\vspace{-0.1cm}
\begin{equation}\label{kmexpansion}
\bm{g}(\bm{x}(t)) = \sum_{i=1}^{n_{d}}\phi_{i}(\bm{x}(t))\,\bm{\upsilon}_{i} = \sum_{i=1}^{n_{d}}\phi_{i}(\bm{x}(0))\,\bm{\upsilon}_{i}\,e^{\lambda_{i}t},
\end{equation}

\vspace{-0.1cm}
\noindent
where $\bm{\upsilon}_{i}\in\mathbb{C}$, \textcolor{black}{$i=1,...,n_{d}$,} are the Koopman modes. The interpretation of (\ref{koopman})--(\ref{kmexpansion}) is straightforward. Instead of focusing on the evolution of the state, $\bm{x}$, one shifts the focus to the observables, $\bm{g}(\bm{x})$. The advantage is that the observables evolve linearly with time, see (\ref{kmexpansion}), without neglecting the nonlinear dynamics of the underlying dynamical system (\ref{eq.1}). The linear representation (\ref{kmexpansion}) is crucial to the proposed method's accuracy and computational efficiency, irrespective of nonlinearities. Note that $\bm{g}(\bm{x})$ can be any continuous function of the state, $\bm{x}$, including the state itself. See, e.g., \cite{Netto2021} for a principled way of selecting these observables. Given a set of observables, it is straightforward to estimate a subset of the Koopman tuples $\{\lambda_{i},\phi_{i},\bm{\upsilon}_{i}\}$. To this end, this work adopts the extended dynamic mode decomposition (EDMD) method \cite{williams2015data}. {\color{black}Following \cite{williams2015data}, ``if the data provided to the EDMD method are generated by a Markov process instead of a deterministic dynamical system, the algorithm approximates the eigenfunctions of the Kolmogorov backward equation, which could be considered as the stochastic Koopman operator.''}

\vspace{-0.2cm}
\section{The Proposed Method}
Let a deterministic power system model be
\vspace{-.1cm}
\begin{equation}
\bm{\dot{x}}=\bm{f}(\bm{x},\bm{y}),\quad \bm{0}=\bm{h}(\bm{x},\bm{y}),
\label{Eq:DAE}
\end{equation}

\vspace{-0.1cm}
\noindent
where {\color{black}$\bm{y}\in\mathbb{R}^{n_{y}}$ denotes} algebraic variables, {\color{black}$\bm{h}:M\to\mathbb{R}^{n_{y}}$} is a nonlinear function, and $\bm{x}$ and $\bm{f}$ are as defined in (\ref{eq.1}). Further, let $\bm{\xi}$ be a random vector following a given probability density function. Now, suppose that $\bm{m}(\bm{\xi})$, a subset of the model parameters\footnote{\color{black}We consider synchronous generators' instead of transmission lines' model parameters because the former directly impact the differential equations.}, is uncertain. To propagate the parameter uncertainty through the system model, consider a set of $n_{mc}$ samples, drawn from a multivariate probability distribution of $\bm{\xi}$, $\{ \bm{\xi}^{(j)} \}^{n_{mc}}_{j=1}$. Then, for each $\bm{\xi}^{(j)}$, $j=1,...,n_{mc}$, one evaluates a modified model given by
\vspace{-0.1cm}
\begin{equation}
\label{stodynamic}
\bm{\dot{x}}=\bm{f}(\bm{x},\bm{y},\bm{m}(\bm{\xi}^{(j)})),\quad \bm{0}=\bm{h}(\bm{x},\bm{y}),
\end{equation}

\vspace{-0.1cm}
\noindent
to obtain $n_{mc}$ trajectories, from which one can quantify the sample mean and the sample variance of the states. Obviously, this MC simulation can be computationally costly for real-time applications in sizeable electric power networks. Now, let us introduce the propagation of parameter uncertainties using the Koopman operator.

\vspace{-0.3cm}
\subsection{Reformulation of the Dynamic Model}
The kernel idea in this letter is to augment (\ref{eq.1}) with $n_{m}$ differential equations \cite{meyers2019koopman}, as follows:
\vspace{-0.1cm}
\begin{equation} \label{modelaugment}
\begin{cases}
\dot{\bm {x}}(t) = \bm {f}(\bm {x}(t),{\bm {m}}(t)), \\
\dot{\bm {m}}(t) = \bm {0}, \\
\end{cases} 
\end{equation}

\vspace{-0.1cm}
\noindent
thereby allowing one to cast the problem of parameter uncertainty propagation into the Koopman operator framework. The dimension of the augmented model (\ref{modelaugment}) is $n_{x}+n_{m}$, where $n_{m}$ is the number of parameters deemed uncertain. Note that generator model parameters are time-invariant, constant values represented by \emph{pseudo-state variables} in (\ref{modelaugment}). Now, we are in the position to act on the parameter space using the Koopman operator formalism. Note that although the model parameters are typically considered time-invariant, as they are here, exceptions do exist, e.g., adaptive control gain in inverter-based resources. Nonetheless, one can still capture these exceptions in (\ref{modelaugment}) as long as ordinary differential equations can describe them; in that case, specifically, $\dot{\bm {m}}(t) = \bm {0}$ would be modified accordingly. This fact demonstrates the flexibility in reformulating the augmented model, though this specific case goes beyond the scope of this letter. 

\subsection{Simulation-Based Data Collection} 
{\color{black}We are now in a position to estimate the Koopman operator. The estimation of the Koopman operator relies exclusively on data, either numerical or experimental. In this letter, we use numerical data obtained from simulations. To this end, we first perturb the initial conditions of the pseudo-states---namely, the parameters ${\bm {m}}(0)$ in \eqref{modelaugment}---at different random values, $\bm{\xi}^{(j)}$, $j=1,\dots,n_t$. More specifically, we adopt a model
\vspace{-.1cm}
\begin{equation}
{\bm{m}}^{(j)}(0)=\bm{m}+{\bm{\xi}^{(j)}}   
\end{equation}

\vspace{-.1cm}
\noindent
to obtain a set $\{ {\bm{m}^{(j)}}(0) \}^{n_t}_{j=1}$, where $n_t$ denotes the number of sampled trajectories. Note that the values of $\bm{m}$ can be obtained from the manufacturer data.} Then, we repeatedly evaluate 
\vspace{-0.1cm}
\begin{equation} 
\label{modelaugmenttrain}
\begin{cases}
\dot{\bm {x}}(t) = \bm {f}(\bm {x}(t),{\bm {m}}(t)), & {\bm{x}({0})}, \\
\dot{\bm {m}}(t) = \bm {0}, & {{\bm{m}^{j}({0})}}, j=1,\dots,n_t, \\
\end{cases} 
\end{equation}

\vspace{-0.1cm}
\noindent
to obtain $n_t$ trajectories of the system states, including the pseudo-states, as the training data. Note that to ensure the training efficiency, $n_t$ should be designed to be a small number while maintaining a faster convergence rate than the MC sampling. Specifically, we generate $\{ \bm{\xi}^{j} \}^{n_t}_{j=1}$ via the Latin hypercube sampling technique for its well-known capability in experiment design. Using the simulated data obtained with the augmented model (\ref{modelaugmenttrain}), we {\color{black}estimate a subset of the Koopman tuples,} $\{\lambda_{i}, \phi_{i}, \bm{\upsilon}_{i}\}$, using the EDMD method \cite{Netto2021, williams2015data}. {\color{black}Note that identifying a Koopman operator-based surrogate model requires the computation of the Moore-Penrose pseudo-inverse of a data matrix. The latter might be time-consuming depending on the matrix dimension and the numerical implementation. Nevertheless, highly efficient implementations of the Moore-Penrose pseudo-inverse are available.}

\vspace{-0.5cm}
\subsection{UQ through Koopman Operator-Based Surrogate Model}
For convenience, define $\bm{x}_{a}^{\top}=[\bm{x}^{\top}\,\bm{m}^{\top}]$. Let us use \eqref{kmexpansion} to mimic the system performances described in \eqref{modelaugment} as a Koopman operator-based surrogate model. Obviously, \eqref{kmexpansion} is in a much simpler functional form than \eqref{modelaugment} to represent a complex dynamical system \cite{lehmberg2020exploring}, such as the dynamic power system considered here. This surrogate allows us to efficiently conduct uncertainty quantification, i.e., $\bm{x}_{ak} \approx \sum _{i=1}^{n_d}\phi _{i} ({\bm{x}_{a0}}) \bm {\upsilon }_{i}\mu _{i}^{k}$, at a large number of parameter values, $\{ {\bm{m}}^{(j)} \}^{n_{mc}}_{j=1}$. {\color{black}Note that $\mu_{i}$ relates to the continuous-time Koopman eigenvalues $\lambda_{i}=\ln\left({\mu_{i}}\right)/\Delta{t}$, where $\Delta{t}$ is the data sampling time.}

To numerically achieve this realization procedure, we simply assign each parameter sample, $\bm{m}^{(j)}$, as the initial conditions to the associated pseudo-states while keeping the initial conditions of the true system states unchanged to get an updated ${\bm{x}_{a0}^{(j)}}$, whose randomness can be further reflected in the \emph{Koopman eigenfunctions} through
$\bm {\phi }({\bm{x}_{a0}^{(j)}})\approx \bm{L} {\bm {g}({\bm{x}_{a0}^{(j)}})}$. The matrix $\bm{L}$ stands for the left eigenvectors of the finite-dimensional approximation to the Koopman operator; refer to \cite{Netto2021} for details. The other part in \eqref{kmexpansion} remains unchanged, and then we have
\vspace{-0.2cm}
\begin{equation}
\label{koopmanevaluation}
{\bm{x}_{ak}^{(j)}} \approx \sum _{i=1}^{n_d}\phi _{i} ({{\bm{x}_{a}}_{0}^{(j)}}) \bm {\upsilon }_{i}\mu _{i}^{k}, \quad j=1,\ldots, n_{mc}.
\end{equation}

\vspace{-0.2cm}
Now, using the set of $\{{{{\bm{x}_{ak}}}^{(j)}}\}_{j=1}^{n_{mc}}$, we can quantify the uncertainties---e.g., the mean, the variance, the probability density function---in the system states at any given time, $k$.

\vspace{-0.3cm}
\section{Simulation Results}
Using the proposed method, we test its performance on the $10$-machine, $39$-bus New England power system with a classic generator model. The system dynamics are triggered by opening Line $15$-$16$. We assume that the parameter values of the inertia for each generator are not well known. We suppose that they follow a Gaussian distribution with the mean being the original manufacturer data and the standard deviation being $10\%$ of the mean value to account for the parameter uncertainties. We use an MC simulation with $10,000$ samples to obtain the benchmark results for comparison. 

\begin{figure}[!t]
\vspace{-.15cm}
\centering 
\includegraphics[scale=0.65]{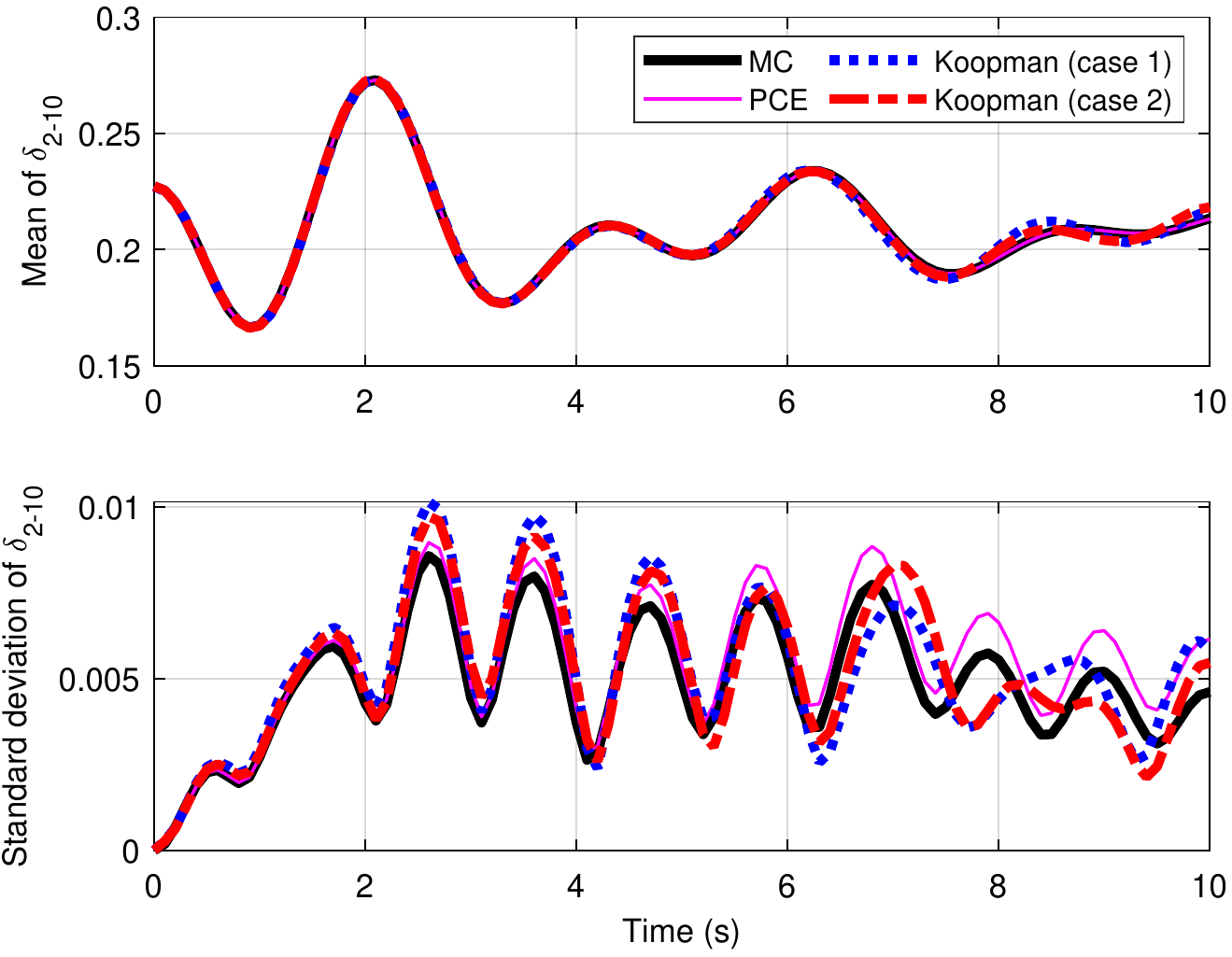}
\vspace{-0.35cm}
\caption{\textcolor{black}{Sample mean and standard deviation of $\delta_{2-10}$ obtained with MC simulation, PCE-based, and Koopman operator-based methods under Gaussian distribution.}}
\label{fig:loop1}
\end{figure}

\begin{figure}[!t]
\vspace{-.3cm}
\centering 
\includegraphics[scale=0.65]{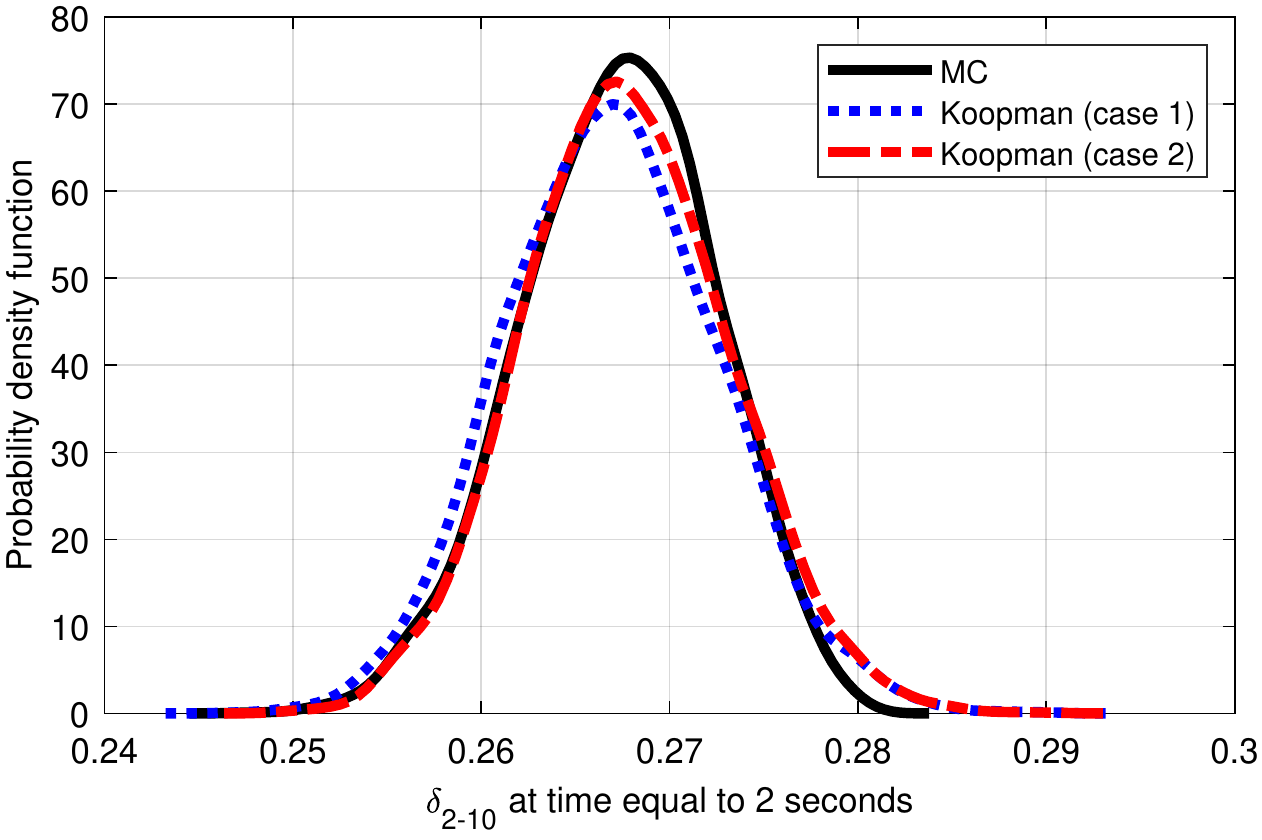}
\vspace{-0.35cm}
\caption{\textcolor{black}{Probability density function of $\delta_{2-10}$ obtained with MC simulation and the Koopman operator-based methods.}}
\label{fig:loop2}
\end{figure}

For the Koopman method, we set $n_{t}=75$, and we select the quantity of interest as the rotor angle of Generator $2$ with respect to that of Generator $10$, denoted as $\delta_{2-10}$, as an example. Note that the choice of the observables for the Koopman operator is an open research topic; therefore, we demonstrate two test cases with different observables. For the first case, we use the second-order multivariate \emph{Hermite polynomials}. For the second case, in addition to the Hermite polynomials, we further introduce a cosine function and a sine function for each true state variable separately. The evolution of their means and standard deviations are depicted in Fig. \ref{fig:loop1}, which shows that, under different observables, their means are quite accurate. Regarding the variance, although small differences are obtained during the first $5$ s, these values increase as time evolves. This makes sense because the errors can accumulate over time \cite{xu2018propagating}. \textcolor{black}{This is precisely what is observed when executing the polynomial-chaos-expansion (PCE) method based on the sparse-grid rule \cite{xiu2010numerical}, whose computing time amounts to $32$ seconds. A similar computing time for the Koopman operator method is recorded. However, while the Koopman operator method has been applied with some success to coherency identification, stability assessment and modal analysis, among others, it still calls for further research. Indeed, the Koopman surrogate approach can not only serve as an alternative of the PCE method in UQ, but it can also help us to better deal with power system uncertainties.} Also, as observed in Fig. \ref{fig:loop2}, which depicts the probability density function of $\delta_{2-10}$ at $t=2$ s, the Koopman surrogate has the capability of accurately representing the full probability density of the system state at a given time. Compared with the MC simulations, which, as indicated in Table \ref{cputime57}, take nearly $0.5$ hour to complete, the Koopman surrogate using Hermite polynomials takes only $0.5$ minute, hence achieving a speedup of more than $50\times$ while maintaining a good accuracy. \textcolor{black}{Note that parallel computing is directly applicable to the training and the partial realization of the Koopman surrogate, resulting in a significant improvement of the computational efficiency of the method.} In addition, these simulations demonstrate the flexibility of using different observables in the Koopman approximation. \textcolor{black}{Note that by adequately tuning the observable functions \cite{Netto2021}, the Koopman surrogate still has the potential to be further improved in long-term dynamic simulations and its commuting efficiency for larger-dimensional systems.}

\begin{figure}[!t]
\centering 
\includegraphics[scale=0.64]{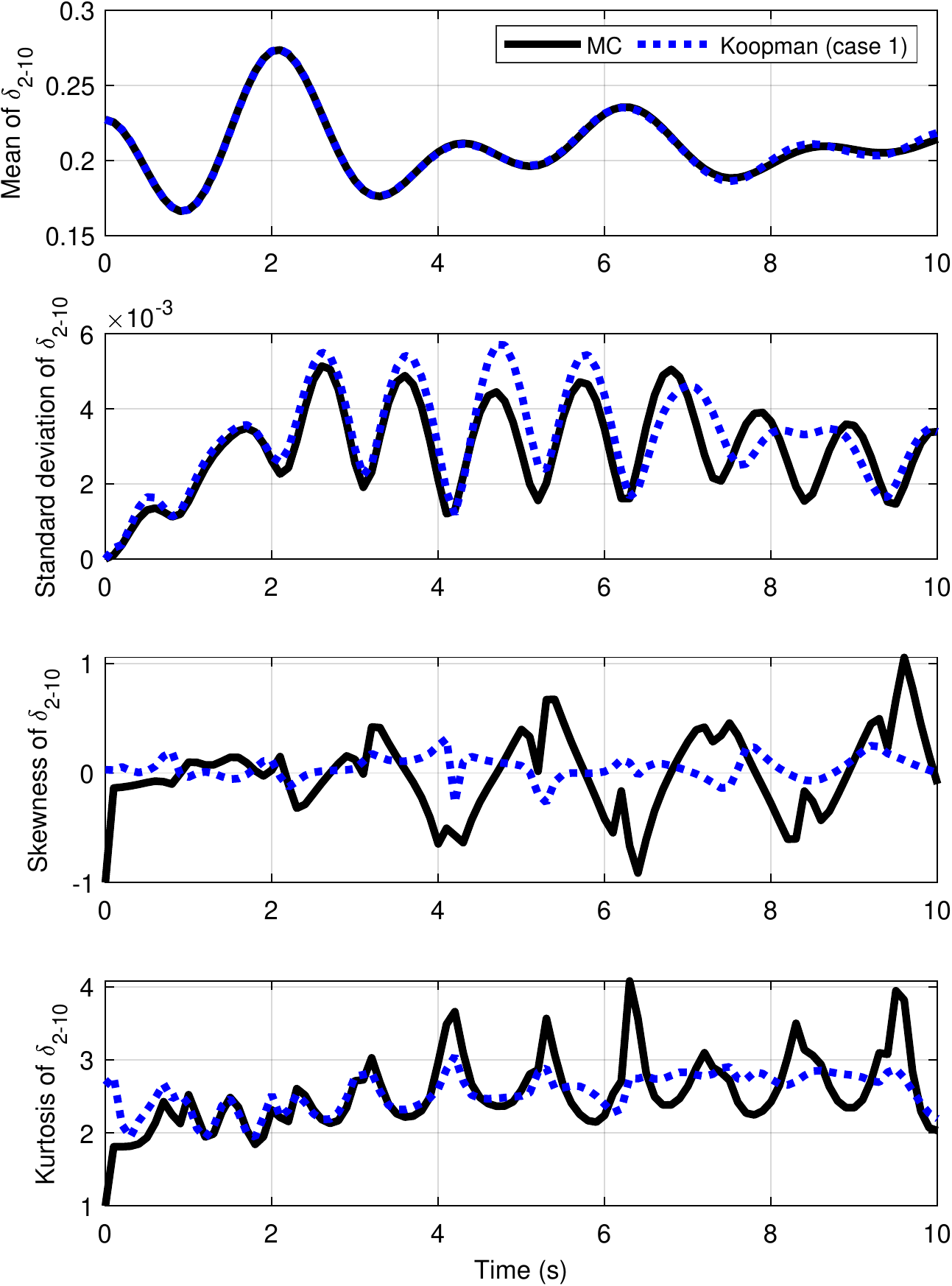}
\vspace{-0.35cm}
\caption{\textcolor{black}{Sample mean, standard deviation, skewness, and kurtosis of $\delta_{2-10}$ obtained with MC simulation and the Koopman operator-based method under Uniform distribution.}}
\label{fig:loop3}
\end{figure}

\textcolor{black}{The proposed method is also able to deal with non-Gaussian uncertainties. Indeed, once the Koopman surrogate is trained, it can be directly evaluated by processing non-Gaussian distributed samples, $\{ {\bm{m}}^{(j)} \}^{n_{mc}}_{j=1}$, to propagate uncertainties. Considering power system applications, let us assume that the parameter values of the inertia for each synchronous generator follows a uniform probability distribution with $10\%$ errors. The other settings remain unchanged. From Fig. \ref{fig:loop3}, we can see that the Koopman method works well in approximating the mean and the variance under an uniform distribution. However, when it comes to higher moments, such as the skewness and the kurtosis as shown in Fig. \ref{fig:loop3}, the Koopman method does not provide accurate results. Therefore, improving the performance of the Koopman method in higher-order moments deserves further exploration.}

\begin{table}[!t]
\centering
\caption{CPU time: MC simulation and Koopman operator-based method}
\vspace{-.15cm}
\begin{tabular}{l c c}
\hline
\textbf{Method} & \textbf{MC} & \textbf{Koopman: Training / Realization / Total} \\ \hline
\textbf{CPU time} & $1627.28$ s & $12.4$ / $15.31$ / $30.83$ s \\
\hline
\end{tabular}
\label{cputime57}
\end{table}

\vspace{-0.2cm}
\section{Conclusions}
In this letter, we propose a Koopman surrogate method for propagating uncertainties in power system dynamic simulations that achieve good performance in terms of accuracy and computational efficiency.

\bibliographystyle{IEEEtran}

\end{document}